\documentstyle[eqsecnum,multicol,aps,prl,epsf]{revtex}

\date{\today}
\draft
\tighten
\begin{document}
\def\sqr#1#2{{\vcenter{\hrule height.3pt
      \hbox{\vrule width.3pt height#2pt  \kern#1pt
         \vrule width.3pt}  \hrule height.3pt}}}
\def\square{\mathchoice{\sqr67\,}{\sqr67\,}\sqr{3}{3.5}\sqr{3}{3.5}}
\def\today{\ifcase\month\or
  January\or February\or March\or April\or May\or June\or July\or
  August\or September\or October\or November\or December\fi
  \space\number\day, \number\year}

\def\Bbb{\bf}



\newcommand {\be}{\begin{equation}}
\newcommand {\ee}{\end{equation}}
\newcommand {\bea}{\begin{array}}
\newcommand {\cl}{\centerline}
\newcommand {\eea}{\end{array}}
\renewcommand {\thefootnote}{\fnsymbol{footnote}}

\def\nc{noncommutative }
\def\com{commutative }
\def\ncy{noncommutativity }
\def\repr{representation }
\def\reps{representations }
\def \simlt{\stackrel{<}{{}_\sim}}
\def \simgt{\stackrel{>}{{}_\sim}}



\title{Discrete Symmetries (C,P,T) in Noncommutative Field Theories} 

\author{M.M. Sheikh-Jabbari}

\address {The Abdus Salam International Center for Theoretical Physics, 
Strada Costiera 11,Trieste, Italy\\
E-mail:jabbari@ictp.trieste.it 
}
\maketitle


\begin{abstract} 
In this paper we study the invariance of the \nc gauge theories under C, P
and T transformations. For the \nc space (when only the spatial part of
$\theta$ is non-zero) we show that NCQED is Parity invariant. In addition,
we show that under charge conjugation the theory on \nc $R^4_{\theta}$ is
transformed to the theory on $R^4_{-\theta}$, so NCQED is a CP violating
theory. The theory remains invariant under time reversal if, together
with proper changes in fields, we also change $\theta$ by  $-\theta$.
Hence altogether NCQED is CPT invariant. Moreover we show that the CPT
invariance holds for general \nc space-time.
\end{abstract}

\pacs{11.15.-q, 11.30.Er, 11.25.Sq, 12.90.+b 
{\qquad} {\qquad} {\qquad} {\qquad} {\qquad} {\qquad}
{\qquad} {\qquad}  ICTP-2000-04}
\vspace*{0.1cm}

\begin{multicols} {2}


\section{Introduction}

Recently it has been shown that the noncommutative spaces arise naturally
when one studies the perturbative string theory in the presence of 
D-branes with non-zero B-field background, i.e. the low energy 
worldvolume theory on such branes is a \nc supersymmetric gauge theory(for 
a review of the field see \cite{SWN}). 

Besides the string theory arguments, the \nc field theories by themselves
are very interesting. Generally, \nc version of a field theory is obtained by
replacing the product of the fields appearing in the action by the
*-product:

\be
(f*g)(x)=exp({i\over 2}\theta_{\mu\nu}\partial_{x_{\mu}}\partial_{y_{\nu}}) 
f(x)g(y)|_{x=y},
\ee
where $f$ and $g$ are two arbitrary functions, which we assume to be
infinitely differentiable. The "Moyal Bracket" of two functions is
$$\{f,g\}_{M.B.}=f*g-g*f.$$
It is apparent that if we choose $f$ and $g$ to be the coordinates
themselves we find
\be\label{NC}
\{x_{\mu},x_{\nu}\}=i\theta_{\mu\nu},
\ee
and this is why these spaces are called noncommutative. Moreover,
consistently we assume the derivatives to act trivially on this space:
\be
\{x_{\mu},\partial_{\nu}\}=-\eta_{\mu\nu}\;\;\;\,
\;\;\;\{\partial_{\mu},\partial_{\nu}\}=0.
\ee

Because of the nature of the *-product, the \nc field theories
for the slowly varying fields or low energies ($\theta E^2\simlt 1$), at {\it classical} level,
effectively reduce to their \com version. However, this is only the classical result and quantum corrections will show
the effects of $\theta$ even at low energies \cite{Shir}.
Since the derivatives are commuting after rewriting the \nc fields and
their action in terms of the Fourier modes we find a \com field theory in
the momentum space, and this field theory has unfamiliar momentum
dependent interactions\cite{Sh2}. 
In this way we find a tool to study these theories perturbatively, like
the usual \com field theories.

It has been shown that the \nc version of $\phi^4$ theory in 4 dimensions
is two loop renormalizable \cite{{Shir},{Aref}}, moreover it is shown
that the \ncy parameter, $\theta$, {\it does not receive quantum corrections}.

The pure \nc U(1) theory has been discussed and shown to be one loop
renormalizable. The one loop beta function for
\nc U(1) is negative (and hence the theory is asymptotically free).
The interesting result is that this one loop beta function {\it is
not $\theta$ dependent} \cite{{Sh2},{MS},{KW}}.  
However, it is not clear whether this property remains at higher loops.
It is believed that all of these one loop properties are a
consequence of the fact that the planar degrees of freedom of \nc theories
is the same as a \com theory \cite{BS}.
The question of the renormalizability has also been addressed for \nc QED
(\nc U(1)+ fermions)\cite{Haya}.

In this paper we study another interesting question about \nc theories
regarding their behaviour under discrete symmetries. 
Since in the noncommutative spaces we have missed the Lorentz symmetry, 
discrete symmetries and in particular the CPT invariance, in the context of
noncommutative geometry in general, are usually non-trivial questions. 
This question has been very briefly discussed in \cite{Haya}.
Hence, first we should build the \nc version of QED, NCQED. We show that
there are two distinct
choices for the fermion representations.  We will show that these two are
related by charge conjugation, so we may call them positively or
negatively charged representations. We will give more intuitive
explanations for these representations. 
In section 3, we study the behaviour of our theory under discrete
symmetries. 
In this section we show the explicit calculations for the cases with
$\theta_{0i}=0$ ($x^0$ is the time coordinate) and we only present the
results for the non-zero $\theta_{0i}$ case in the last part of this
section. For the $\theta_{0i}=0$ cases,
we show that our theory, NCQED, is parity invariant, with the
usual transformation of the fields; and studying the charge conjugation
transformations we show that the the NCQED {is not} C-invariant and in
order to make the theory invariant besides the usual field transformations
we should also change $\theta$ by $-\theta$. In addition we show that the
same $\theta$ changing is needed for time reversal invariance. So,
although our theory is CP violating, it is CPT invariant.
For {\it the general $\theta$} we show that though C , P and T are
broken, the whole theory is again CPT invariant.
The last section is devoted to conclusions and remarks.

\vskip .5cm
\section{Building the NCQED}
{\it i)Pure Gauge theory}

The action for the pure gauge theory is
\be\label{AA}
S={1\over 4\pi}\int F_{\mu\nu}*F^{\mu\nu}d^4x=
{1\over 4\pi}\int F_{\mu\nu}F^{\mu\nu}d^4x
\ee
with 
\be
F_{\mu\nu}=\partial_{[\mu}A_{\nu]}+ig \{A_{\mu},A_{\nu}\}.
\ee
In the above $g$ is the gauge coupling constant. Let us consider the
following transformations
\be\bea{cc}\label{UA}
A_{\mu}\rightarrow A'_{\mu}= 
U(x)*A_{\mu}*U^{-1}(x)+\\ {i\over g}U(x)*\partial_{\mu}U^{-1}(x),\\
U(x)=exp*(i\lambda),\;\;\;\;\ U^{-1}(x)=exp*(-i\lambda),
\eea\ee
where
\be\bea{cc}
exp*(i\lambda(x))\equiv 1+i\lambda-{1\over 2} \lambda*\lambda-{i\over 3!}
\lambda*\lambda*\lambda+...\\
U(x)*U^{-1}(x)=1.
\eea 
\ee
Under the above transformations
\be
F_{\mu\nu}\rightarrow F'_{\mu\nu}=U(x)*F_{\mu\nu}*U^{-1}(x).
\ee
Hence due to the cyclic property of integration over space (see
Appendix A) the action is invariant under (\ref{UA}). The above argument
can be trivially generalized to U(N) cases by letting $A$ and $\lambda$
take values in the related algebra. However, in this paper we only consider 
the U(1) case.

{\it ii)Fermionic Part}

In order to write down the fermionic part of the action with the \nc U(1)
symmetry mentioned above, first we need to find the proper
"fundamental" representations of the \nc U(1) group. There are 
two distinct choices for that:

a) The \repr with 
\be\left\{\bea{cc}
\psi_+(x)\rightarrow \psi'_+(x)=U(x)*\psi_+(x),\\
{\bar\psi}_+(x)\rightarrow {\bar\psi}'_+(x)={\bar\psi}_+(x)*U^{-1}(x),
\eea\right.
\ee

and

b) the other with
\be\left\{\bea{cc}
\psi_-(x)\rightarrow \psi'_-(x)=\psi_-(x)*U^{-1}(x), \\
{\bar\psi}_-(x)\rightarrow {\bar\psi}'_-(x)=U(x)*{\bar\psi}_-(x).
\eea\right.
\ee
We will show that this two type of fermions are related by a charge
conjugation transformation. The next step is finding
a "covariant" derivative, $D_{\mu}$. For these two types of
fermions we have different "covariant" derivatives 

a)\be
D_+^{\mu}\psi_+(x)\equiv
\partial^{\mu}\psi_+(x)-ig\psi_+(x)*A^{\mu}(x),
\;\;{\rm and}
\ee
   
and
b)\be
D_-^{\mu}\psi_-(x) \equiv
\partial^{\mu}\psi_-(x)+igA^{\mu}(x)*\psi_-(x).
\ee
One can show that with each  of the covariant derivatives defined above
(with the proper fermionic \repr ), the action
\be
S=\int d^4x \bar\psi*(i\gamma^{\mu}D_{\mu}\psi -m\psi)
\ee
is invariant under the \nc U(1) transformations.

\section{P, C and T Invariance}

Having the form of the action we are ready to study the P, C, and T
symmetries.
For the sake of certainty up to the last part of this section we consider
the \nc spaces, i.e. $\theta_{0i}=0$, and in the last paragraph we discuss
the non-zero $\theta_{0i}$ and the most general \nc space-time.

\vskip.3cm
{\bf{\it Parity}}

Under the parity, $x_i\rightarrow -x_i$, the $\theta$ parameter is not
changed (see (\ref{NC})). It is straightforward to show that 
for parity transformation given by:
\be
\left\{ \bea{cc}
A_0\rightarrow A_0 \\
A_i\rightarrow -A_i \\
\psi(x)\rightarrow \gamma^0\psi\\
x_i\rightarrow -x_i,
\eea\right.
\ee
the NCQED action is invariant for both of the fermionic choices.

\vskip .3cm
{\bf\it Charge Conjugation}

Let us first study the pure \nc U(1) case. Under the usual charge
conjugation, C-, transformations,
\be\label{CA}
A_{\mu}\rightarrow -A_{\mu},
\ee
(\ref{AA}) is not invariant, because the first term in the $F$ will change
the sign but the second term, $\{A_{\mu}, A_{\nu}\}$, will remain the
same. To make the theory C invariant we note that the Moyal bracket
changes the sign if together with (\ref{CA}) we also change $\theta$ by 
\be\label{Cth}
\theta\rightarrow -\theta.
\ee

The above $\theta$ transformation has an intuitive explanation. 
As discussed in \cite{{Dip},{Sh2}}, the gauge symmetry of \nc U(1) is an
infinite dimensional algebra which its Cartan subalgebra (the zero
momentum sector) is a U(1) leading to a photon like state, and all
the other gauge particles look like {\it dipoles} under this U(1), whose 
dipole moment is proportional to the $\theta$.  
So, in this picture we feel the necessity of (\ref{Cth}), under charge
conjugation.

Hence, the \nc U(1) theory with parameter $\theta$ is mapped into another 
\nc U(1) theory with $-\theta$. 

Now, we should consider the fermionic part. 
Since the kinetic part of the
fermionic action is unchanged, we take the usual C-transformations:
\be\label{FC}
\left\{\bea{cc}
\psi \rightarrow i\gamma^0\gamma^2 {\bar\psi}^T=-i\gamma^2\psi^*\\
{\bar\psi} \rightarrow i\psi^T\gamma^2\gamma^0.
\eea\right.
\ee

Let us first discuss the fermions in $+$ \repr (type a) fermions). 
Under the above transformations, {\it without changing $\theta$}
\be\bea{cc}
\int d^4x \bar\psi*(i\gamma^{\mu}A_{\mu}(x)*\psi)\rightarrow\\
-\int d^4x \bar\psi*(i\gamma^{\mu}\psi*A_{\mu}(x)),
\eea\ee
we see that this is exactly the form of interaction term for the type b)
fermions. In other words, the types a) and b) fermions are charge conjugate
of each other. Let us consider the $\theta$ transformation too. By using
roles given in Appendix A, we see that forms of the interaction term
for these two types of fermions are related by (\ref{Cth}),
which means that (\ref{FC}) together with (\ref{Cth}) and (\ref{CA})
give  proper charge conjugation transformations, a discrete symmetry of NCQED. 

\vskip .3cm
{\bf{\it Time Reversal}}

First we consider the pure \nc U(1) and then we study fermions.
Under the time reversal, in order to keep the kinetic part of our gauge
field action, $A_{\mu}$ should transform as
\be\label{TA}
\left\{ \bea{cc}
A_0\rightarrow A_0 \\
A_i\rightarrow -A_i.
\eea\right.
\ee
 
Now let us look at the terms with Moyal brackets. Since time reversal
operator involves a complex conjugation, for any two real fields, $f$ and
$g$ we have
\be
f*g|_{\theta}\rightarrow f_T*g_T|_{-\theta}=g_T*f_T|_{\theta},
\ee
where $f_T$ and $g_T$ show the time reversed $f$ and $g$ respectively,
then we have
\be
\{f,g\}_{M.B.}\rightarrow -\{f_T,g_T\}_{M.B.}.
\ee
Since $A_{\mu}$'s are real fields,
\be
ig \{A^{\mu},A^{\nu}\}\rightarrow ig \{A_T^{\mu},A_T^{\nu}\},
\ee    
and
\be\bea{cc}
F_{0i}=\partial_{[0}A_{i]}+ ig \{A_{0},A_{i}\}\rightarrow \\
\partial_{[0}A_{i]}-ig \{A_{0},A_{i}\},\\
F_{ij}=\partial_{[i}A_{j]}+ ig \{A_{i},A_{j}\}\rightarrow \\
F_{ij}=\partial_{[i}A_{j]}-ig \{A_{i},A_{i}\},
\eea\ee
the only way to make the theory invariant under time reversal is
changing $\theta$ as well as $A_{\mu}$:
\be\label{Tth}
\theta\rightarrow -\theta.
\ee
So (\ref{TA}) together with (\ref{Tth}) give the proper time reversal
transformations.

{\it The Fermionic Part}

Since the kinetic term is quadratic in fields, $\psi$'s should obey the
usual time reversal transformations:
\be\label{FT}
\left\{\bea{cc}
\psi \rightarrow i\gamma^1\gamma^3 {\psi} \\
{\bar\psi} \rightarrow i{\bar\psi}\gamma^1\gamma^3.
\eea\right.
\ee

As for the interaction term, for the sake of certainty let us consider the
type a) case, {\it without changing $\theta$}. We find  
\be\bea{cc}
\int d^4x \bar\psi*(i\gamma^{\mu}A_{\mu}(x)*\psi)\rightarrow \\
\int d^4x {\bar\psi}_T*(i\gamma^{*\mu}A^T_{\mu}(x)*\psi_T)|_{-\theta},
\eea\ee
where $\gamma^{*\mu}$ is the complex conjugate of $\gamma^{\mu}$.
Replacing $\psi_T$ and $A_T$ from (\ref{FT}) and (\ref{TA}), we obtain
\be\bea{cc}
\int d^4x {\bar\psi}_T*(i\gamma^{*\mu}A^T_{\mu}(x)*\psi_T)|_{-\theta}= \\
-\int d^4x \bar\psi*(i\gamma^{\mu}A_{\mu}(x)*\psi)|_{-\theta},
\eea\ee
which is exactly the interaction term for type b) fermions. As we see,
in order to make the NCQED time reversal invariant, we should consider
(\ref{Tth}), (\ref{FT}) and (\ref{TA}) together.

\vskip .5cm
{\bf{\it CPT}}

Now that we have studied P, C, and T, it is interesting to consider the CP
and CPT too. As we showed parity transformations remain the same as the
\com version, however each of C and T involves an extra
$\theta\rightarrow-\theta$. So altogether the NCQED (with parameter
$\theta$) is CP violating, i.e. , it maps the theory into NCQED with
$-\theta$, and the theory is CPT invariant.
We should note that although our system is not manifestly Lorentz invariant CPT, as an
accidental symmetry, remains valid. 

\vskip .3cm
{\bf{\it Non-zero $\theta_{0i}$ and general $\theta_{\mu\nu}$}}

Although a well-defined Hamiltonian for non-zero $\theta_{0i}$ cases is not found yet
and hence the quantum theory for these cases is not understood in the same sense as $\theta_{ij}$ case, 
one can formally study the discrete symmetries for these cases.
As it is readily seen from (1.2) under parity the $\theta_{0i}$ components
should be replaced with $-\theta_{0i}$, and we can show that the (3.1)
transformations together with this $\theta$ change is the symmetry of
NCQED.

For the charge conjugation to make the theory invariant the change in
$\theta$ parameter, (3.3), should be extended to $\theta_{0i}$ components
too.

It is straightforward to check that the time reversal invariance is
achieved if $\theta_{0i}$ are unchanged while the $\theta_{ij}$ components
should be transformed by (\ref{Tth}). Having in mind that time reversal 
involves a complex conjugation, this result is expected from (1.2).
Hence for general $\theta_{\mu\nu}$ the theory remains CPT invariant,
although the theory violates P, C and T.

\section{Conclusions and Remarks} 

In this paper we have reviewed the \nc gauge theory and their gauge
symmetry and shown that fermions can be added in two distinct fundamental
\reps of the gauge group. We have shown that these two \reps are
related by charge conjugation, so we called them positive or negative
representations. 

Studying the discrete symmetries for the $\theta_{0i}=0$ cases, we
have shown that NCQED is parity invariant under the usual (\com) field
transformations.
For C and T transformations we showed that besides the usual field
transformations we need an extra $\theta\rightarrow-\theta$
transformation. In other words, NCQED with $\theta$ is charge conjugated (or 
time reversed) of NCQED with $-\theta$. Therefore, despite being Lorentz non-invariant, 
in this case NCQED is CT invariant, and hence CPT invariant. In other words P and CPT is an accidental
symmetry of the system.

For the general $\theta_{\mu\nu}$, we discussed that P, C and T
invariance are broken, however the the theory is again CPT invariant.

Noncommutative gauge theories seem to provide a very good framework for
the {\it CP violating} models, which are of great importance in particle
physics phenomenology. The advantage of these theories is that the beta function is not $\theta$
dependent and furthermore $\theta$ does not receive quantum corrections.
Therefore the amount of CP violation is completely under control.

{\bf Acknowledgements} \vskip.3cm
I would like to thank D.Demir and Y. Farzan for fruitful discussions. I would also
like to thank D. Ployakov for reading the manuscript.

This research was partly supported by the EC contract no. ERBFMRX-CT 96-0090.
\vskip .3cm
{\large{\bf Appendix:} Some Useful Identities in
*-product calculus}  

Let $f, g$ be two arbitrary functions on \nc $R^d$:
$$
f(x)=\int f(k) e^{ik.x}d^dk, \;\;\;\;
g(x)=\int g(k) e^{ik.x}d^dk.
$$
Then
$$
(f*g)(x)=\int f(k) g(l)e^{-ik\theta l/2} e^{i(k+l).x}d^dkd^dl,
$$
where $k\theta l=k^{\mu}\theta_{\mu\nu}l^{\nu}$. From the above relation
it is straightforward to see:

1) $g*f=f*g|_{\theta\rightarrow -\theta}$, and hence
$\{f,g\}_{M.B.}=f*g|_{\theta} -f*g|_{-\theta}$.

2) $\int (f*g) (x) d^dx=\int (g*f) (x) d^dx=\int fg (x) d^dx$.

3) If we denote  complex conjugation by $c.c.$, then

$(f*g)^{c.c.}=g^{c.c.}*f^{c.c.}$. 

If $h$ is another arbitrary function:

4) $(f*g)*h=f*(g*h)\equiv f*g*h$.

5)$\int (f*g*h)(x) d^dx=\int (h*f*g)(x) d^dx=\int (g*h*f)(x) d^dx$.

6) $(f*g*h)|_{\theta}= (h*g*f)|_{-\theta}$.

In other words the integration on the space coordinates, $x$, has the
cyclic
property, and it has all the properties of the $Tr$ in the matrix calculus.

From 2) we learn that the kinetic part of the actions (which are quadratic
in fields) is the same as their commutative version. So the free field
propagators in \com and \nc spaces are the same.

%
%

\end{multicols}
\end{document}